\documentclass[apjl,chicago,twocolumn,numberedappendix,twocolappendix]{emulateapj}
\usepackage{graphicx}
\usepackage{mathrsfs}
\usepackage[intlimits,centertags]{amsmath}
\usepackage{amssymb,amsfonts}
\usepackage{enumerate}
\usepackage[pdftex]{hyperref}
\usepackage[x11names]{xcolor}
\usepackage{booktabs}
\usepackage{mathpazo}
\usepackage{pifont}
\usepackage{multirow}

\hypersetup{pdftitle={Large- and Medium-Scale Anisotropies in the Arrival Directions of Cosmic Rays observed with KASCADE-Grande},
pdfsubject={Large- and Medium-Scale Anisotropies in the Arrival Directions of Cosmic Rays observed with KASCADE-Grande},
pdfauthor={Markus Ahlers},
pdfstartview={FitH},
colorlinks=true,
bookmarksopen=false,
bookmarksnumbered=false,
bookmarksopenlevel=0,
linkcolor=Blue1!60!black,
citecolor=Green1!50!black,
urlcolor=Blue1!70!black
}

\begin{document}

\title{Large- and Medium-Scale Anisotropies in the Arrival Directions of Cosmic Rays\\ observed with KASCADE-Grande}

\author{Markus Ahlers}
\shortauthors{Markus Ahlers}
\affiliation{Niels Bohr International Academy \& Discovery Center, Niels Bohr Institute,\\ University of Copenhagen, Blegdamsvej 17, DK-2100 Copenhagen, Denmark}

\begin{abstract}
We search for anisotropies in the arrival directions of cosmic rays observed by the KASCADE-Grande air shower experiment. The analysis is based on public data of about 23.7 million events with reconstructed primary energies above 1 PeV. We apply a novel maximum-likelihood reconstruction method for the cosmic ray anisotropy, that compensates for spurious anisotropies induced by local detector effects. We find no evidence for a large-scale dipole anisotropy in the data, consistent with official results based on the conventional East--West derivative method. On the other hand, a subset of cosmic rays with median energy of 33 PeV shows strong evidence for a medium-scale feature with an angular diameter of 40 degrees. After accounting for the look-elsewhere effect, the post-trial significance of this medium-scale feature is at the level of $4\sigma$.
\end{abstract}

\keywords{cosmic rays --- methods: data analysis}

\maketitle

\section{Introduction}

Cosmic rays (CRs) experience deflections by Galactic and extragalactic magnetic fields before their arrival on Earth. The spatial variation of these magnetic fields in strength and orientation scrambles the particles' arrival direction and time. Combined with the limited energy resolution and livetime of CR observatories, these effects can explain the continuity of the flux of CRs and the mostly isotropic distribution of their arrival directions. However, some CR experiments have achieved the necessary level of statistics to be able to infer weak anisotropies in the arrival directions that reach a per-mille level at TeV--PeV energies and even a percent level above the ankle~\citep{DiSciascio:2014jwa,Ahlers:2016rox,Deligny:2018blo}.

The size and strength of the residual anisotropy are controlled by the spatial and temporal distribution of CR sources and magnetic field configurations. The dipole anisotropy observed below 2~PeV can be understood in terms of the presence of nearby sources -- presumably supernova remnants -- and anisotropic diffusion in local magnetic fields~\citep{Ahlers:2016njd}. This large-scale anisotropy could induce the observed medium- and small-scale features by CR streaming through local random magnetic field configurations~\citep{Giacinti:2011mz,Ahlers:2013ima,Ahlers:2015dwa}. Extragalactic CRs above 8~EeV show a significant large-scale dipole feature with an amplitude of a few percent. This observation can be interpreted as an excess from an extragalactic source distribution, distorted by magnetic fields~\citep{Aab:2017tyv}. 

So far, no significant CR anisotropies have been detected in the intermediate range from 2~PeV to 8~EeV. It has been argued that the best-fit dipole phases inferred from data in this energy range exhibit a smooth transition between adjacent energy bins and could indicate a continuous transition between source populations~\citep{Deligny:2018blo}. However, the significance of this observation compared to random fluctuations is debatable. In any case, a robust identification of anisotropies would provide valuable data to decipher the transition between Galactic and extragalactic CR sources.

In this Letter we search for anisotropies in the arrival directions of CRs observed with the KASCADE-Grande air shower experiment~\citep{Haungs:2018xpw}. The analysis is based on public data provided by the KASCADE Cosmic Ray Data Center (KCDC) and uses a novel maximum-likelihood reconstruction method introduced in \cite{Ahlers:2016njl}, that we outline in the following section. We first discuss the presence of a dipole anisotropy in the data and compare our results to those derived via the conventional East--West derivative method~\citep{Bonino:2011nx}. We then study -- for the first time -- the presence of medium-scale anisotropies in the KASCADE-Grande data.

\section{Cosmic-Ray Anisotropy Reconstruction}

Due to the diffusive dispersion of arrival times, the flux of CRs can be considered as continuous over the livetime of ground-based observatories. In a fixed energy range, we can express the flux (units of ${\rm cm}^{-2}\,{\rm s}^{-1}\,{\rm sr}^{-1}$) as
\begin{equation}\label{eq:phi}
  \phi(\alpha,\delta) = \phi^{\rm iso}I(\alpha,\delta)\,,
\end{equation} 
where $\phi^{\rm iso}$ is the angular-averaged isotropic flux level and $I(\alpha,\delta)$ is the relative intensity in terms of right ascension $\alpha$ and declination $\delta$. Cosmic ray diffusion predicts that the {\it anisotropy} $\delta I = I-1$ is subdominant, $|\delta I|\ll1$.

In the local reference system of a ground-based observatory the arrival directions of CRs are uniquely characterized by their azimuth angle $\varphi$, zenith angle $\theta$, and local sidereal time $t$.  A unit vector ${\bf n}'(\varphi, \theta)$ in the local horizontal coordinate system is related to the corresponding unit vector ${\bf n}(\alpha, \delta)$ in the celestial equatorial coordinate system via a coordinate transformation ${\bf n}={\bf R}(t)\cdot{\bf n}'$. The rotation matrix ${\bf R}$ depends on local sidereal time $t$ and the geographic latitude $\Phi$ of the observatory; see, {\it e.g.}, \cite{Ahlers:2016njl}. At any time, the observatory's field of view is limited by a maximum zenith angle $\theta_{\rm max}$. Over the course of many sidereal days, the observatory then covers a time-integrated field of view in the equatorial coordinate system that is characterized by the declination band, $\delta_{\rm min}<\delta<\delta_{\rm max}$, with $\delta_{\rm min} = {\rm max}(-90^\circ,\Phi-\theta_{\rm max})$ and $\delta_{\rm max} = {\rm min}(90^\circ,\Phi+\theta_{\rm max})$.

\begin{figure}[t]\centering
\includegraphics[width=\linewidth]{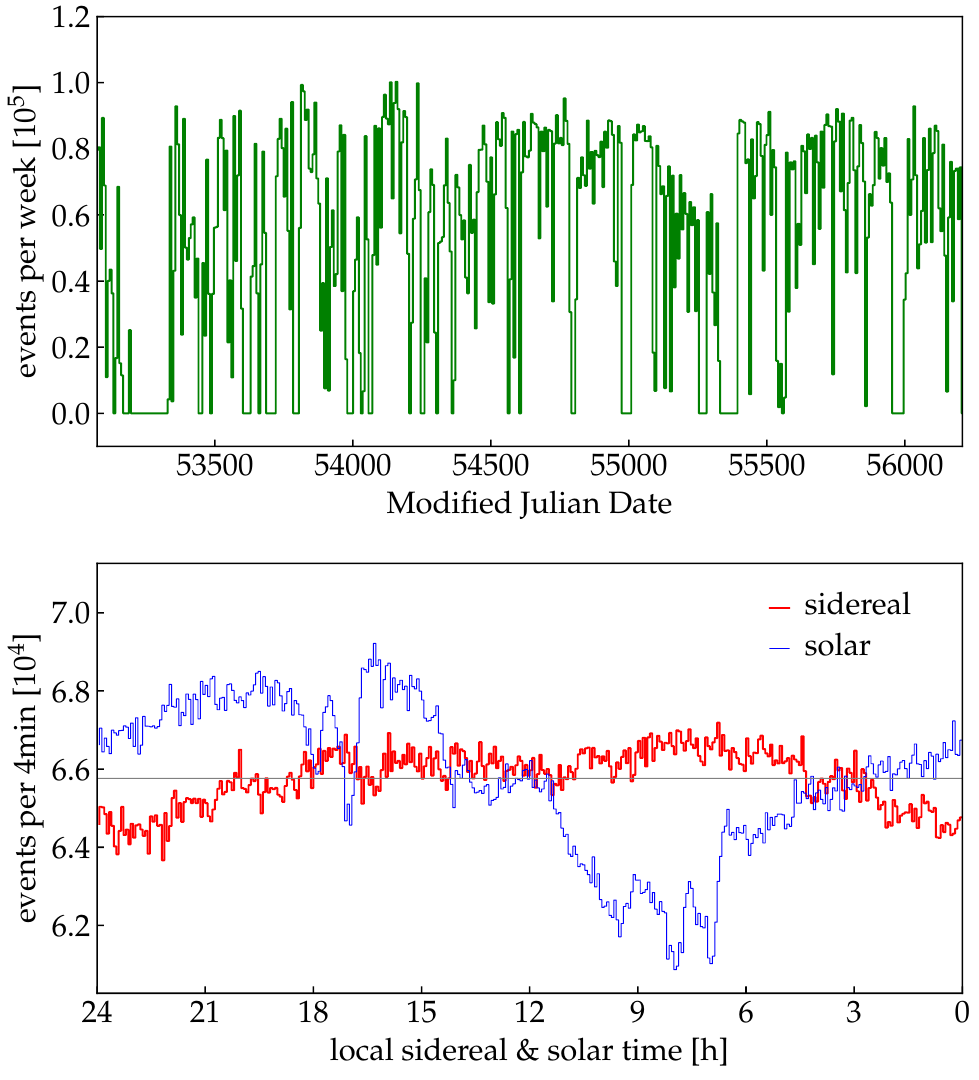}
\caption[]{Distribution of KASCADE-Grande events with $N_{\rm ch}\geq10^{5.2}$ over Modified Julian Days (top), solar time (bottom; thin blue line), and local sidereal time (bottom; thick red line).}\label{fig1}
\end{figure}

We will assume in the following that the detector exposure $\mathcal{E}$ per solid angle and sidereal time $t$ accumulated over many sidereal days can be expressed as a product of its angular-integrated exposure $E$ per sidereal time (units of ${\rm cm}^2\, {\rm sr}$) and relative acceptance $\mathcal{A}$ (units of ${\rm sr}^{-1}$ and normalized as $\int{\rm d}\Omega \mathcal{A}(\Omega)=1$):
\begin{equation}\label{eq:E}
  \mathcal{E}(t,\varphi,\theta) \simeq E(t)\mathcal{A}(\varphi,\theta)\,.
\end{equation}
The same assumption is also implicit in CR background estimations by direct integration~\citep{Atkins:2003ep} or time-scrambling~\citep{Alexandreas1993}. Note that the accumulation of data into sidereal bins tends to average out variations in the relative acceptance that are out of phase with the length of one sidereal day. 

To simplify calculations on the local and celestial spheres, the sky is binned into pixels of equal area $\Delta\Omega$ using the {\tt HEALPix} parameterization of the unit sphere~\citep{Gorski:2004by}. We follow the convention in~\cite{Ahlers:2016njl} and use {\it roman} indices for pixels in the local sky map and {\it fraktur} indices for pixels in the celestial sky map. Time bins are indicated by {\it greek} indices. For instance, the data observed at a fixed sidereal time bin $\tau$ can be described in terms of the observation in the local horizontal sky with bin $i$ as $n_{\tau i}$ or transformed into the celestial sky map with bin $\mathfrak{a}$ as $n_{\tau \mathfrak{a}}$. 

The number of CRs expected from within the solid angle $\Delta\Omega$ in the direction ${\bf n}'(\varphi_i,\theta_i)$ and within a sidereal time interval $\Delta t$ with a central value $t_\tau$ is
\begin{equation}\label{eq:mu}
  \mu_{\tau i} \simeq I_{\tau i}\mathcal{N}_\tau\mathcal{A}_{i}\,,
\end{equation}
where $\mathcal{N}_\tau\equiv\Delta t\phi^{\rm iso}{E}(t_\tau)$ gives the expected number of isotropic background events in sidereal time bin $\tau$. The quantity $\mathcal{A}_i \equiv \Delta\Omega\mathcal{A}(\varphi_i,\theta_i)$ is the relative acceptance in the local bin $i$, and $I_{\tau i}\equiv {I}({\bf R}(t_\tau){\bf n}'(\varphi_i,\theta_i))$ is the corresponding relative intensity observed in the time bin $\tau$. Given $\mu_{\tau i}$, the likelihood of observing $n_{\tau i}$ CRs is given by the product of Poisson probabilities
\begin{equation}\label{eq:LH}
  \mathcal{L}(n|I,\mathcal{N},\mathcal{A}) =
  \prod_{\tau i}\frac{(\mu_{\tau i})^{n_{\tau i}}e^{-\mu_{\tau i}}}{n_{\tau i}!}\,.
\end{equation}
The maximum-likelihood (max-$\mathcal{L}$) combination of parameters $(I^\star,\mathcal{N}^\star,\mathcal{A}^\star)$ for given data $n$ can be inferred via an iterative reconstruction method as outlined and validated in \cite{Ahlers:2016njl} and \cite{Ahlers:2018qsm}.

The likelihood-based anisotropy reconstruction has several advantages compared to the conventional East--West derivative method; see the Appendix. The max-$\mathcal{L}$ method {\it i)} compensates for detector effects without prior assumptions on the local angular acceptance, {\it ii)} delivers a two-dimensional representation of the anisotropy, {\it iii)} allows to combine data from different observatories in a joint analysis, {\it e.g.}~\cite{Aartsen:2018ppz}, and {\it iv)} provides a direct statistical measure to quantify the significance of anisotropies at various angular scales.

\begin{table*}[t]\renewcommand{\arraystretch}{1.3}\renewcommand{\tabcolsep}{3pt}
\centering
\caption[]{Reconstructed dipole anisotropy}\label{tab1}
\begin{minipage}[t]{0.99\linewidth}
\begin{ruledtabular}
\begin{tabular}{cccc|cc|cc|ccccc}
\multicolumn{4}{c}{}&\multicolumn{2}{c}{East--West (official)\tablenotemark{\ding{169}}}&\multicolumn{2}{c}{East--West (this work)}&\multicolumn{5}{c}{max-$\mathcal{L}$ (this work)\tablenotemark{\ding{171}}}\\
data&$E_{\rm med}$\tablenotemark{\ding{168}}&$N_{\rm ch}$-range&$N_{\rm tot}$&$A$ [$10^{-3}$]&$\alpha$ [${}^\circ$]&$A$ [$10^{-3}$]&$\alpha$ [${}^\circ$]&$A$ [$10^{-3}$]&$\alpha$ [${}^\circ$]&$\lambda$&$p$-value&{$A_{90}$~[$10^{-3}$]}\\
\hline
sidereal&\multirow{2}{*}{--}&\multirow{2}{*}{$\geq10^{5.2}$}&\multirow{2}{*}{$23,674,844$}&$2.8\pm0.8$&$227\pm17$&${2.9\pm1.3}$&${228\pm26}$&$2.1\pm0.9$&$266\pm24$&5.52&$0.063$&{$3.7$}\\
solar&&&&$1.5\pm0.8$&$359\pm32$&${2.7\pm1.3}$&${337\pm29}$&$1.1\pm0.9$&$357\pm40$&1.61&$0.45$&{$2.5$}\\
\hline
bin 1 &$2.7$~PeV&$[10^{5.2},10^{5.6})$&$17,443,774$&$2.6\pm1.0$&$225\pm22$&${3.4\pm1.5}$&${218\pm26}$&$2.1\pm1.0$&$243\pm27$&$4.49$&$0.11$&{$3.7$}\\
bin 2 &$6.1$~PeV&$[10^{5.6},10^{6.4})$&$6,084,275$&$2.9\pm1.6$&$227\pm30$&${1.9\pm2.7}$&${281\pm82}$&$3.3\pm1.8$&$314\pm31$&$3.46$&$0.18$&{$6.0$}\\
bin 3 &$33$~PeV&$\geq10^{6.4}$&$146,795$&$12\pm 9$&$254\pm42$&${24\pm18}$&${240\pm42}$&$9\pm11$&$299\pm77$&$0.57$&$0.75$&{$28$}
\end{tabular}
\end{ruledtabular}
\centering{${}^{\text{\ding{168}}}$ based on \cite{Chiavassa:2015jbg}\hspace{0.5cm}${}^{\text{\ding{169}}}$ results presented in \cite{Apel:2019afz}\hspace{0.5cm}${}^{\text{\ding{171}}}$ method introduced in \cite{Ahlers:2018qsm}}
\end{minipage}
\end{table*}

\section{Analysis of KASCADE-Grande Data}

The KASCADE-Grande experiment located in Karls\-ruhe, Germany ($49^\circ\!\!.\,1$\,N, $8^\circ\!\!.\,4$\,E) is a CR observatory collecting charged particles created in extended CR air showers. The footprint of the CR shower observed on the ground level allows us to reconstruct the arrival direction of CRs. The reconstructed number of charged particles in the shower, $N_{\rm ch}$, serves as a proxy of the initial CR energy. The data used in this analysis were collected between March 2004 and October 2012 and are available via KCDC~\citep{Haungs:2018xpw} as one of the preselected data products: {\tt ReducedData-GRANDE\_runs\_4775-7398\_HDF5}. The arrival direction of events in this data set is limited to zenith angles below $40^\circ$. For a comparison to previous anisotropy studies by the KASCADE-Grande Collaboration~\citep{Chiavassa:2015jbg,Apel:2019afz} we select high-energy events with $N_{\rm ch}\geq10^{5.2}$ and bin the data into three $N_{\rm ch}$ bins that are listed in the third column of Table~\ref{tab1}. The median energy of these $N_{\rm ch}$ bins has been inferred from Monte Carlo simulation in \cite{Chiavassa:2015jbg} and is shown in the second column. The data distributions in terms of Modified Julian Date as well as solar and local sidereal time are shown in Fig.~\ref{fig1}.

\subsection{Large-scale Anisotropy}

We will first study the presence of a dipole anisotropy in the KASCADE-Grande data using the max-$\mathcal{L}$ method presented in \cite{Ahlers:2018qsm}. It is important to realize that this method does not allow to reconstruct anisotropies that are azimuthally symmetric in the equatorial coordinate system (see Appendix A in \cite{Ahlers:2018qsm}). The {\it reconstructable} dipole anisotropy is therefore of the form 
\begin{equation}\label{eq:Idipole}
\delta I_{\rm dipole}(\alpha,\delta) = d_x\cos\alpha\cos\delta + d_y\sin\alpha\cos\delta\,.
\end{equation}
With this ansatz for the relative intensity, we can reconstruct the maximum combination $(d^\star_x,d^\star_y,\mathcal{N}^\star,\mathcal{A}^\star)$ of  Eq.~(\ref{eq:LH}) using an iterative method. After a few iteration steps (about $20$ in this analysis), the max-$\mathcal{L}$ ratio between the best-fit dipole anisotropy and the null hypothesis,
\begin{equation}
\lambda = 2\ln\frac{\mathcal{L}(n|d_x^\star,d_y^\star,\mathcal{N}^\star_\tau,\mathcal{A}^\star_i)}{\mathcal{L}(n|0,0,\mathcal{N}^{(0)}_\tau,\mathcal{A}^{(0)}_i)}\,,
\end{equation}
allows us to estimate the significance of the dipole anisotropy. Data following the null hypothesis have a distribution in $\lambda$ that follows a two-dimensional $\chi^2$-distribution~\citep{Wilks:1938dza}. The $p$-value of the observed data, {\it i.e.}, the probability of a false-positive identification of the dipole anisotropy, is then simply given by $p= e^{-\lambda/2}$. In addition, the best-fit values of $\mathcal{N}^\star$ and $\mathcal{A}^\star$ allow us to estimate the uncertainties of the best-fit dipole anisotropy (see \cite{Ahlers:2018qsm} for details).

The last five columns of Table~\ref{tab1} show our results on the dipole anisotropy based on the max-$\mathcal{L}$ method for the combined data binned in local sidereal time (first row) and solar time (second rows) as well as the three $N_{\rm ch}$ bins in sidereal time (last three rows). The best-fit range including the 68\% confidence level (C.L.) is expressed in terms of the amplitude $A_1$ and phase $\phi_1$ of the dipole projected onto the equatorial plane. We also indicate the test-statistic value $\lambda$ and the corresponding $p$-value. We find no evidence for a dipole anisotropy in the individual data sets. The last column shows the 90\% C.L.~upper limit on the dipole amplitude.

For a better comparison with previous KASCADE-Grande analyses~\citep{Chiavassa:2015jbg,Apel:2019afz} (columns 5 \& 6) we also study the dipole anisotropy with the East--West derivative method (columns 7 \& 8); see the Appendix. The best-fit amplitudes and their standard deviations inferred with this method are somewhat larger than official results. This seems to be related to different values of the effective right ascension step $\Delta \alpha$ in Eq.~(\ref{eq:EWdipole}); whereas~\cite{Apel:2019afz} choose $20^\circ$ we derive values between $12^\circ$ and $13^\circ$ based on Eq.~(\ref{eq:dalphaestimator}). We have checked from reconstructions of Monte Carlo data that this expression provides an unbiased estimator of $\Delta \alpha$. One can notice that the max-$\mathcal{L}$ reconstruction is more precise, {\it i.e.},~the standard deviation on the dipole amplitude is smaller than $\sqrt{2/N_{\rm tot}}/\Delta\alpha$ expected from the East--West derivative method (see Appendix C in \cite{Ahlers:2018qsm}).

Note that the analysis of \cite{Apel:2019afz} applied an additional quality cut to the data, discarding events having the largest particle density measured by station number 15. This subset of events shows a strong nonuniform azimuthal distribution in the local coordinate system. The max-$\mathcal{L}$ method does not require this quality cut, since the reconstruction does not rely on symmetries of the local angular acceptance.

\begin{figure*}[t]\centering
\includegraphics[width=\linewidth]{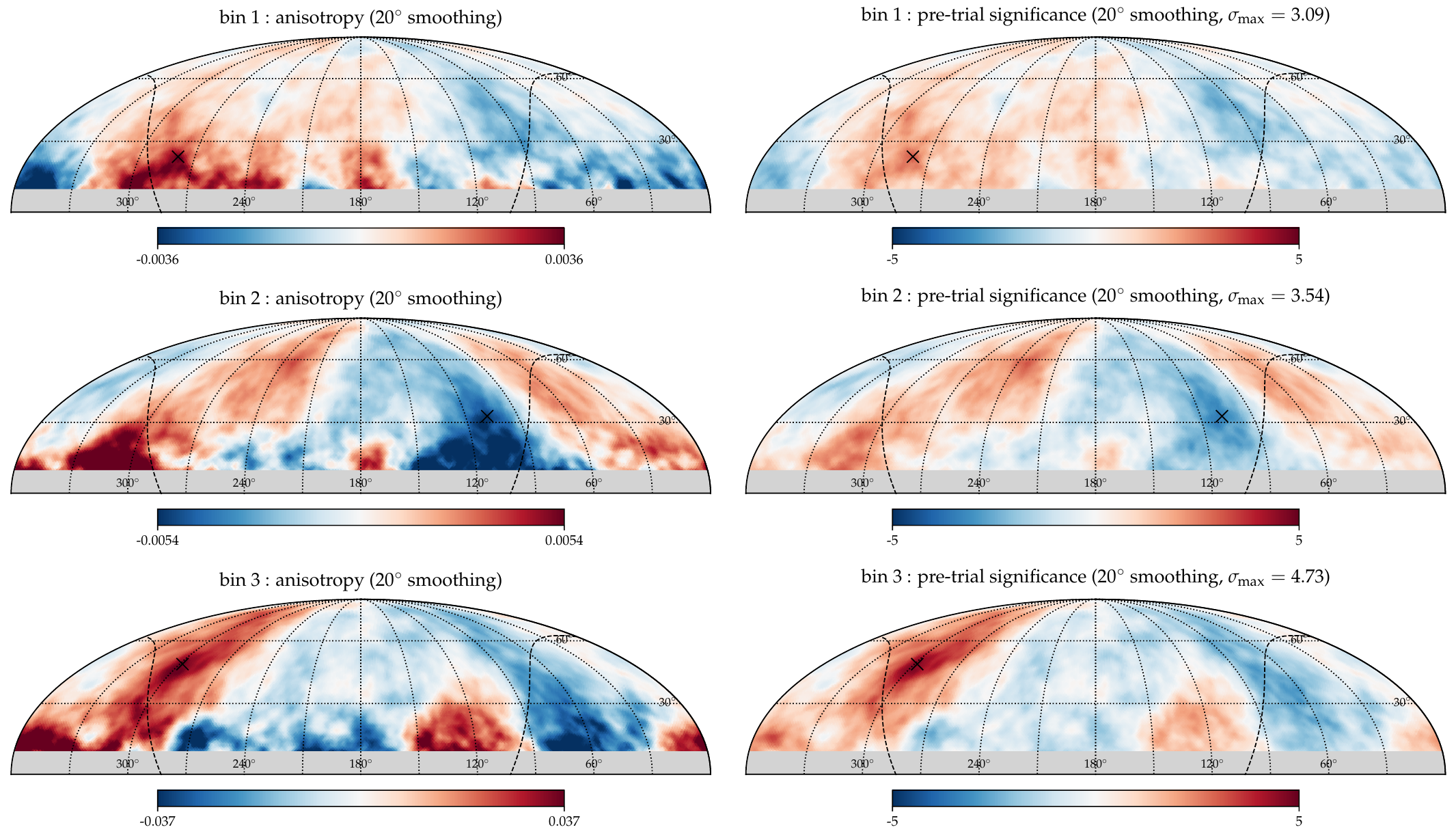}
\caption[]{Mollweide projections in equatorial coordinates of the reconstructed anisotropy (left) and pre-trial significance (right) for the three $N_{\rm ch}$ bins listed in Table~\ref{tab1}. We show the results for a top-hat smoothing radius of $20^\circ$. The gray-shaded area indicates the unobservable part of the celestial sphere. The dashed line indicates the projection of the Galactic plane. The values of pre-trial significance are shown in units of standard deviations and indicated by negative values for deficits. The location of maximum pre-trial significance is indicated by the symbol $\boldsymbol\times$.}\label{fig2}
\end{figure*}

\subsection{Medium-scale Anisotropy}

The likelihood-based anisotropy reconstruction allows to study the presence of anisotropies at arbitrary angular scales by a bin-wise fit of $\delta I$ in the equatorial coordinate system. The likelihood is again maximized by an iterative reconstruction presented by \cite{Ahlers:2016njl}. Similar to \cite{Ahlers:2018qsm}, we increase the stability of the iterative reconstruction by smoothing the data with a Gaussian symmetric beam with full width half maximum of $2^\circ$. To extract the presence of medium-scale anisotropies we smooth the resulting anisotropy and event numbers by a top-hat kernel with radius of $20^\circ$. This corresponds to the sum of events and expectation values over the set $\mathcal{D}_\mathfrak{a}$ of data bins within $20^\circ$ off a central bin $\mathfrak{a}$ in the equatorial coordinate system:
\begin{align}\label{eq:rebinning1}
\widetilde{n}_{\mathfrak{a}} &= \sum_{\mathfrak{b}\in\mathcal{D}_{\mathfrak{a}}}\sum_{\tau}n_{\tau \mathfrak{b}}\,,\\\label{eq:rebinning2}
  \widetilde{\mu}_{\mathfrak{a}} &=  \sum_{\mathfrak{b}\in\mathcal{D}_{\mathfrak{a}}}\sum_{\tau}\mathcal{A}^\star_{\tau\mathfrak{b}}\mathcal{N}^\star_\tau I^\star_{\mathfrak{b}}\,,\\\label{eq:rebinning3}
  \widetilde{\mu}^{\,\rm bg}_{\mathfrak{a}} &= \sum_{\mathfrak{b}\in\mathcal{D}_{\mathfrak{a}}}\sum_{\tau}\mathcal{A}^\star_{\tau\mathfrak{b}}\mathcal{N}^\star_\tau I^{\,\rm bg}_{\mathfrak{b}}\,.
\end{align}
In the absence of strong large-scale anisotropies, the isotropic background level is simply taken as $I^{\,\rm bg}=1$, but can in general take on any form that is considered as the background level. With these definitions we can express the smoothed anisotropy as 
\begin{equation}\label{eq:rebinnedI}
\delta\widetilde{I}_\mathfrak{a} = \widetilde{\mu}_\mathfrak{a}/\widetilde{\mu}^{\,\rm bg}_\mathfrak{a} - 1\,.
\end{equation}
The left panels of Fig.~\ref{fig2} show the reconstructed anisotropy in the three energy bins with excesses and deficits indicated by red and blue colors, respectively. The dashed line indicates the projection of the Galactic plane onto the celestial sphere. 

With the expectation values of Eqs.~(\ref{eq:rebinning1})--(\ref{eq:rebinning3}) we can also define a smoothed significance map as
\begin{equation}\label{eq:rebinnedsig}
  \widetilde{S}_{\mathfrak{a}} \equiv \sqrt{2\big(-\widetilde{\mu}_{\mathfrak{a}}+\widetilde{\mu}^{\,\rm bg}_{\mathfrak{a}} + \widetilde{n}_\mathfrak{a}\log(1+\delta\widetilde{I}_\mathfrak{a})\big)}\,.
\end{equation}
This expression represents the statistical weight of the anisotropy $\delta\widetilde{I}_\mathfrak{a}$ in each celestial (sliding) bin $\mathfrak{a}$. For sufficiently small smoothing scales, $\widetilde{S}_\mathfrak{a}^{\,2}$ can be interpreted as the bin-by-bin maximum-likelihood ratio of the hypothesis $I^\star_\mathfrak{a}$ compared to the null hypothesis $I^{\,\rm bg}_\mathfrak{a}=1$. Again, the test statistic of data under the null hypothesis is following a one-dimensional $\chi^2$-distribution and, in that case, $\widetilde{S}_{\mathfrak{a}}$ corresponds to the significance in units of Gaussian standard deviations~\citep{Wilks:1938dza}.

The right panels of Figure~\ref{fig2} show the pre-trial significance (\ref{eq:rebinnedsig}) of the anisotropy. We follow the standard convention to indicate the significance of deficits by negative values. The symbol ${\boldsymbol\times}$ indicates the location of maximum significance. Whereas the first two bins do not show strong evidence of CR anisotropies, the last bin shows a local excess at the level of about $4.7\sigma$. However, the significance of this excess needs to be corrected for trials. We follow the same procedure as in \cite{Ahlers:2018qsm} to estimate the effective number of trials as $N_{\rm trial} \simeq \Delta\Omega_{\rm FOV}/\Delta\Omega_{\rm bin}$, where $\Delta\Omega_{\rm FOV}$ is the size of the observatory's time-integrated field of view and $\Delta\Omega_{\rm bin}$ is the effective bin size according to the top-hat smoothing scale. For the $20^\circ$ smoothing radius of the KASCADE-Grande data we obtain $N_{\rm trial}\simeq14.0$. The post-trial $p$-value can then be approximated as 
\begin{equation}\label{eq:trialcorrection}
p_{\rm post} \simeq 1-(1-p)^{N_{\rm trial}}\,.
\end{equation}
Figure~\ref{fig3} shows the post-trial significance map for the third KASCADE-Grande bin in Galactic coordinates. As before, the gray-shaded region indicates the part of the sky that is not observable from the location of the experiment. The dashed circle indicates the $20^\circ$ smoothing radius around the location of the highest post-trial significance of about $4.2\sigma$.

\section{Discussion}

Our analysis does not uncover significant dipole anisotropies in the KASCADE-Grande data, as indicated by the $p$-value in the second-to-last column of Table~\ref{tab1}. This is consistent with official results summarized in \cite{Apel:2019afz} and shown in columns 5 \& 6. The dipole amplitude in solar time induced by the solar Compton--Getting effect~\citep{CG1935} is expected to reach an amplitude of only $4.5\times10^{-4}$~\citep{Ahlers:2016rox}, which is far below the 90\% C.L.~upper limit of about $2.5\times10^{-3}$ inferred by our max-$\mathcal{L}$ analysis (see the last column of Tab.~\ref{tab1}). On the other hand, a sidereal dipole anisotropy has been observed in an analysis of IceTop data at a median energy of $1.6$~PeV at the level of $1.6\times10^{-3}$~\citep{Aartsen:2016ivj}. This is consistent with the best-fit sidereal dipole amplitude observed in the first $N_{\rm ch}$ bin with a median energy of $2.7$~PeV.

\begin{figure}[t]\centering
\includegraphics[width=\linewidth]{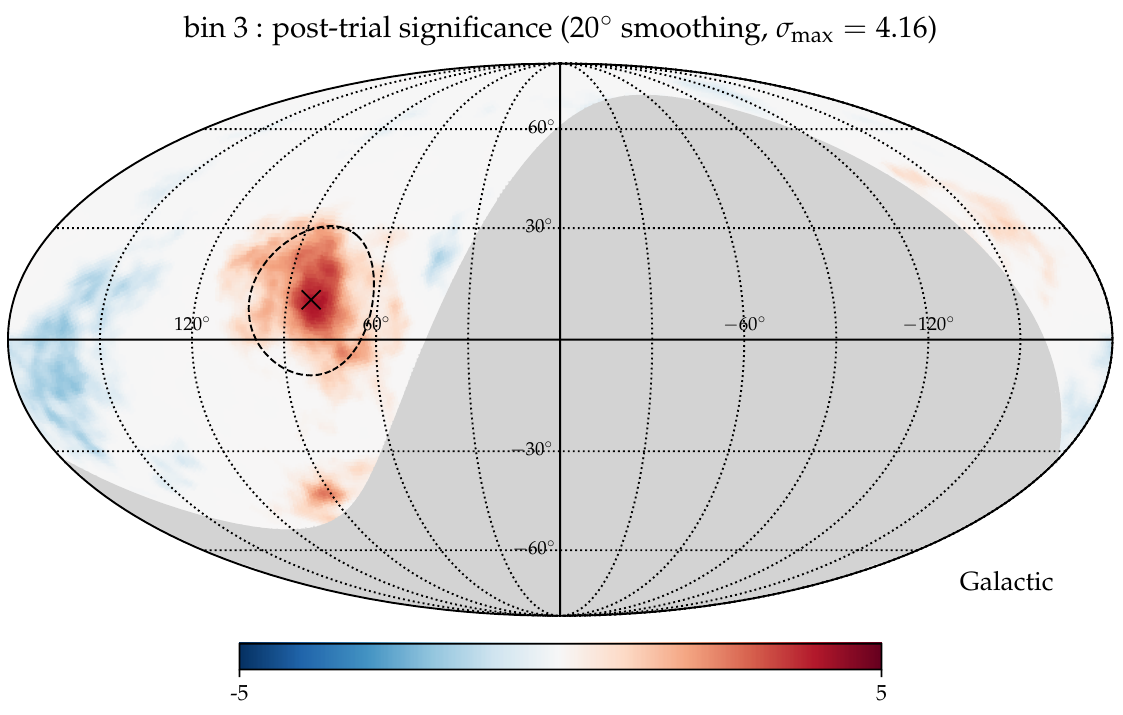}
\caption[]{Mollweide projection in Galactic coordinates of the post-trial significance of $20^\circ$ smoothed anisotropies at 33~PeV (bin 3). We use a trial factor $N_{\rm trials}\simeq 14$ in Eq.~(\ref{eq:trialcorrection}) and show units of Gaussian standard deviations. We indicate the location of the maximum significance by the symbol ${\boldsymbol\times}$ and the $20^\circ$ smoothing radius by a dashed line.}\label{fig3}
\end{figure}

Our analysis finds -- for the first time -- $4\sigma$ evidence for CR anisotropies on angular scales of $40^\circ$ at a level of $3.7\times10^{-2}$ and a median energy of $33$~PeV. The CR flux associated with the excess can be estimated as $E^2\phi_{\rm CR}(E) \simeq 1.7\times10^{-7} {\rm GeV}\, {\rm cm}^{-2}\,{\rm s}^{-1}$. As discussed earlier, the origin of medium-scale anisotropies could be induced by CR streaming in local magnetic fields. The gyroradius of $33$~PeV charged CRs in Galactic magnetic fields is less than 10~pc, and it is therefore not expected that this excess is related to the presence of a local CR source. However, there are two notable exceptions that we highlight in the following.

Neutrons can be produced by CR collisions with gas and reach a decay length of about 300~pc at $33$~PeV. The corresponding anisotropy from local sources would appear fuzzy and distorted due to the variance of the neutron's lifetime and residual magnetic deflections after neutron decay into protons. Interestingly, the smoothing region of the maximal excess shown in Fig.~\ref{fig3} encloses the location of the Cygnus region -- a rich region of gas and star formation in our local Galactic environment.

Another non-diffusive origin of the excess could be a local source of PeV $\gamma$-rays. These $\gamma$-rays would also originate from high-energy CR interactions in the vicinity of their sources. Cosmic ray diffusion before interaction would account for the extended emission. At 33~PeV, the fraction of an isotropic $\gamma$-ray flux in the CR data is below $10^{-3}$, which can be inferred by a search for muon-poor showers~\citep{Apel:2017ocm}. This is marginally consistent with the medium-scale excess at a level of $3.7\times10^{-2}$, if we account for the finite extension of the smoothing region. Diffuse $\gamma$-ray data at GeV--TeV energies would allow to further test this hypothesis~\citep{Abdo:2008if,Ackermann:2012pya,Bartoli:2015era}.

Cosmic ray interactions that yield neutrons and $\gamma$-rays will also be visible in high-energy neutrinos. For instance, if we consider that at least one charged pion is created in the production of a neutron that carries about 25\% of the energy of neutron, the corresponding flux of PeV muon neutrinos is expected to reach a level of $E^2\phi_{\nu_\mu+\bar\nu_\mu}(E) \simeq 1.1\times10^{-8} {\rm GeV}\,{\rm cm}^{-2}\,{\rm s}^{-1}$. This spatially extended emission could be detectable by neutrino observatories like IceCube and ANTARES~\citep{Aartsen:2014cva,Adrian-Martinez:2014wzf,Aartsen:2019epb,Illuminati:2019oag}.

\begin{figure*}[t]\centering
\includegraphics[width=0.95\linewidth]{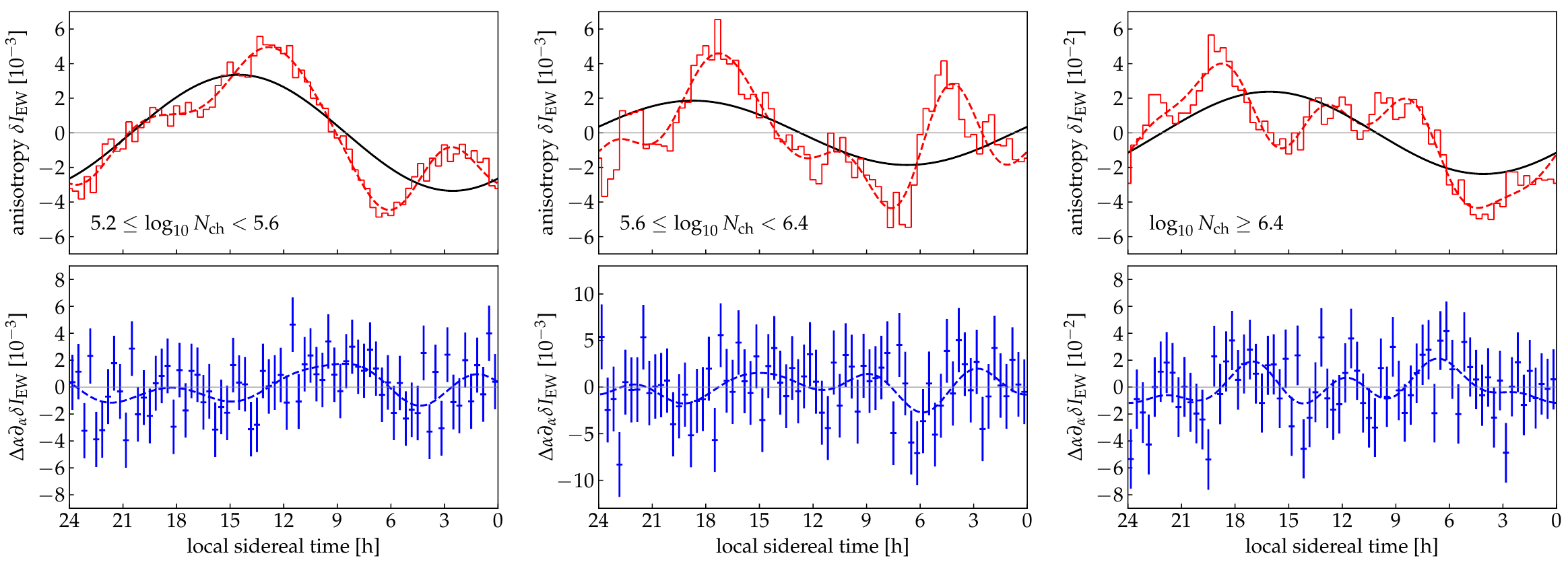}
\caption[]{Reconstruction of the large-scale anisotropy with the East--West derivative method. The lower panels show the differential East--West anisotropy data with the best-fit derivative (including the first five harmonics) indicated as blue dashed lines. The top panels show the corresponding anisotropy data and fit, where the first harmonic is indicated as a black line.}\label{fig4}
\end{figure*}

\section*{Acknowledgements}
I would like to thank the KASCADE-Grande Collaboration for sharing their data via KCDC. In particular, I would like to thank Andrea Chiavassa, Ralph Engel, Andreas Haungs, Donghwa Kang, Dmitriy Kostunin, Markus Roth and J\"urgen Wochele for their feedback on KASCADE-Grande data and previous anisotropy analyses. This work was supported by \textsc{Villum Fonden} under project no.~18994. 

\appendix

\section{East--West Derivative Method}\label{appA}

The East--West (EW) derivative method~\citep{Bonino:2011nx} accounts for variations in the angular acceptance and livetime of the detector by studying the derivative of the relative intensity with respect to right ascension. At each sidereal time $t$ the CR data is divided into two bins, covering the east ($0<\varphi<\pi$) and west ($-\pi<\varphi<0$) sectors in the local coordinate system. The event numbers observed during a short sidereal time interval $\Delta t$ in the east ($+$) and west ($-$) sector can be expressed as
\begin{equation}
N_\pm(t) \simeq \phi^{\rm iso}\Delta t E(t)\int_{0}^{\pi}{\rm d}\varphi\!\!\!\int_0^{\theta_{\rm max}}\!\!{\rm d}\theta\sin\theta\,\mathcal{A}(\pm\varphi,\theta)I(t,\pm\varphi,\theta)\,.
\end{equation}
The EW asymmetry at sidereal time $t$ is then defined as
\begin{equation}\label{eq:EW}
A_{\rm EW}(t) \equiv \frac{N_{+}(t)-N_{-}(t)}{N_{+}(t)+N_{-}(t)}\,.
\end{equation}
We can write the local detector acceptance as $\mathcal{A} = \mathcal{A}_s(1 +\delta \mathcal{J})$, where $\mathcal{A}_s$ is even under EW reflection, $\varphi \to -\varphi$, and $\delta \mathcal{J}$ is odd. For ground-based observatories we expect that $|\delta \mathcal{J}|\ll 1$. To first order in the CR anisotropy $\delta I$ and the asymmetry of the detector acceptance $\delta\mathcal{J}$ we can evaluate the EW derivate as
\begin{equation}
A_{\rm EW}(t) \simeq \langle \delta \mathcal{J} \rangle + \frac{1}{2}\bigg(\langle\delta I(t,\varphi,\theta)\rangle - \langle\delta I(t,-\varphi,\theta)\rangle\bigg)\,,
\end{equation}
where $\langle\cdot\rangle$ denotes the average over the the East sector, $0<\varphi<\pi$, with weight $\mathcal{A}_s$. If we assume that the true anisotropy follows a dipole, we can further reduce this equation to 
\begin{equation}\label{eq:EWdipole}
A_{\rm EW}(t) \simeq \langle \delta \mathcal{J} \rangle + \Delta\alpha\partial_\alpha\delta I(\alpha,0)\,,
\end{equation}
with effective right ascension step size
\begin{equation}\label{eq:dalpha}
\Delta\alpha = \langle\sin\theta\sin\varphi\rangle\,.
\end{equation}
The EW method only allows to study the components of the dipole anisotropy in the equatorial plane, which is a limitation that is also present in the max-$\mathcal{L}$ method. Equations~(\ref{eq:EWdipole}) and (\ref{eq:dalpha}) {\it define} the EW derivative $\partial_\alpha I_{\rm EW}$. It is important to emphasize that, in general, $\partial_\alpha I_{\rm EW} \neq \partial_\alpha I(\alpha,0)$ if the anisotropy deviates from a pure dipole. 

After binning the data into $N_{\rm sid}$ sidereal time bins $\tau$ and $N_{\rm pix}$ celestial bins $i$ one can derive an estimator of the EW asymmetry as
\begin{align}\label{eq:Aestimator}
\widehat{A}_{{\rm EW},\tau} = \bigg(\sum_{i\in\mathcal{D}_+} \!\!n_{\tau i} - \sum_{i\in\mathcal{D}_-}\!\!n_{\tau i}\bigg)\bigg/\sum_i n_{\tau i}\,,
\end{align}
where $\mathcal{D}_\pm$ are the sets of bins in the east ($+$) and west ($-$). The residual EW asymmetry from the detector is independent of sidereal time and can be estimated as
\begin{equation}\label{eq:Jestimator}
\widehat{\langle \delta \mathcal{J}\rangle}  = \frac{1}{N_{\rm sid}}\sum_\tau \widehat{A}_{{\rm EW},\tau}\,.
\end{equation}
The estimator of the effective right ascension step is given by the average
\begin{equation}\label{eq:dalphaestimator}
\widehat{\Delta\alpha} = \frac{1}{2N_{\rm sid}}\sum_\tau\bigg({\Delta\alpha}_{+,\tau}+{\Delta\alpha}_{-,\tau}\bigg)\,,
\end{equation}
where at individual time steps we have
\begin{equation}
{\Delta\alpha}_{\pm,\tau} = \sum_{i\in\mathcal{D}_\pm}\!\!n_{\tau i}\sin\theta_i|\sin\varphi_i|\bigg/\!\!\!\sum_{i\in\mathcal{D}_\pm}\!\!n_{\tau i}\,.
\end{equation}
The leading-order statistical uncertainty in each bin is given by
\begin{equation}
\Delta (\Delta \alpha\partial_\alpha I_{{\rm EW},\tau}) \simeq \bigg(\sum_i n_{\tau i}\bigg)^{-1/2}\,.
\end{equation}
Finally, the lower panels of Fig.~\ref{fig4} show the binned EW derivative for the three KASCADE-Grande bins. The dashed blue line in the lower plots represents the best fit to the data including the first five harmonics. The derivative data and best fit can be converted via Eq.~(\ref{eq:dalphaestimator}) to the corresponding EW anisotropy $\delta I_{\rm EW}$, as shown in the upper panels of Fig.~\ref{fig4}. The black line indicates the best-fit dipole component with best-fit values and standard deviations listed in columns 7 \& 8 of Table~\ref{tab1}. Our analysis reproduces the official results in~\cite{Apel:2019afz} within statistical uncertainties.

\newpage
\bibliographystyle{apj}

\end{document}